\documentclass[aps,twocolumn,pra,superscriptaddress,notitlepage,floatfix]{revtex4-1}
\pdfoutput=1
\usepackage{graphicx,color,xcolor}
\usepackage{marvosym}
\usepackage{amsmath,amssymb}
\usepackage{ulem}

\usepackage{graphicx,amsmath,graphics}

\usepackage[colorinlistoftodos]{todonotes}
\usepackage[colorlinks=true, allcolors=blue]{hyperref}

\makeatletter
\def\subsubsection{\@startsection{subsubsection}{3}{10pt}{-1.25ex plus -1ex minus -.1ex}{0ex plus 0ex}{\normalsize\bf}}
\def\paragraph{\@startsection{paragraph}{4}{10pt}{-1.25ex plus -1ex minus -.1ex}{0ex plus 0ex}{\normalsize\textit}}
\renewcommand\@biblabel[1]{#1}
\renewcommand\@makefntext[1]%
{\noindent\makebox[0pt][r]{\@thefnmark\,}#1}
\DeclareRobustCommand\onlinecite{\@onlinecite}
\def\@onlinecite#1{\begingroup\let\@cite\NAT@citenum\citealp{#1}\endgroup}
\def\tagform@#1{\maketag@@@{\ignorespaces#1\unskip\@@italiccorr}}
\let\orgtheequation\theequation
\def\theequation{(\orgtheequation)}
\makeatother

\newcommand{\ry}{Rydberg }

\begin{document}

\title{Ultralong-range \ry bi-molecules}

  \author{Rosario Gonz\'alez-F\'erez}
\email{rogonzal@ugr.es}%
\affiliation{Instituto Carlos I de F\'{\i}sica Te\'orica y Computacional,
and Departamento de F\'{\i}sica At\'omica, Molecular y Nuclear,
 Universidad de Granada, 18071 Granada, Spain} 
 \affiliation{ITAMP, Harvard-Smithsonian Center for Astrophysics, Cambridge, Massachusetts 02138, USA} 
\author{Janine Shertzer}
\affiliation{ITAMP, Harvard-Smithsonian Center for Astrophysics, Cambridge, Massachusetts 02138, USA} 
\affiliation{Department of Physics, College of the Holy Cross, Worcester, Massachusetts 01610, USA}
\author{H. R. Sadeghpour}
\affiliation{ITAMP, Harvard-Smithsonian Center for Astrophysics, Cambridge, Massachusetts 02138, USA} 
\date{\today}

\begin{abstract} 
We predict that ultralong-range \ry bi-molecules form in collisions between polar molecules in cold and ultracold settings. The collision of $\Lambda$-doublet nitric oxide (NO) with long-lived \ry NO($nf$, $ng$) molecules forms ultralong-range \ry bi-molecules with GHz energies and kilo-Debye permanent electric dipole moments. The Hamiltonian includes both the anisotropic charge-molecular dipole interaction and the electron-NO scattering.  The rotational constant for the \ry bi-molecules is in the  
MHz range, allowing for microwave spectroscopy of rotational transitions in \ry bi-molecules. Considerable orientation of NO dipole can be achieved. The \ry molecules described here hold promise for studies of a special class of long-range bi-molecular interactions.
\end{abstract}

%\pacs{}}

\maketitle

Rydberg spectroscopy is one of the most precise techniques in the AMO physics toolkit to probe core interactions, perform cavity quantum electrodynamics~\cite{Haroche2013}, diagnose environmental influence in laboratory and astrophysical 
gases~\cite{Allard1982,Szudy1996}, engineer strong quantum correlations, and  transmit quantum entanglement over macroscopic distances~\cite{Saffmann2010}. In cold and pristine ultracold AMO traps, \ry interactions with surrounding atoms or molecules  may, under certain conditions, lead to the formation of special long-range Rydberg molecules~\cite{greene00,bendkowsky09}. Such \ry molecules have proven to be fertile grounds for investigating novel few-body phenomena, many-body quantum 
correlations~\cite{Schmidt2016,camargo2018,sous2020}, and exotic ionic interactions~\cite{meinert2018,Hummel2020}. To date, ultracold \ry excitations have been performed in a target atom- usually an alkali metal or alkaline earth 
atom~\cite{Shaffer2018,camargo2018}. 

Ultracold polar molecules have accessible internal degrees of freedom, which can be engineered for simulations of chemical reactions~\cite{ChemPhysChemRev2016,cold_mol,julienne2012}, 
many-body interactions~\cite{zoller2007,DeMarco853}, and information processing in the quantum limit~\cite{yelin2006,zoller2006}. As with \ry systems, polar molecules can possess large dipole moments, making them amenable to control and manipulation.
In addition, interactions involving molecules are fundamental to chemical synthesis, whose studies are in vogue in cold and ultracold traps as densities increase~\cite{Ospelkaus853,Hu1111,Narevicius2019,Cornish2020}. In most cold and ultracold settings, atom-molecule or molecule-molecule collisions often are exothermic, releasing energy leading to losses from the trap. Elaborate schemes to control and shield trapped molecules from such deleterious encounters are 
proposed~\cite{Gorshkov2008,karman2018,karman2019}.

Here, we investigate a particular class of bi-molecular collision, facilitated by \ry excitation in a $\Lambda$-doublet molecule(NO), leading to formation of exotic \ry bi-molecules, whose electronic and rovibrational properties can be readily controlled. Because the $\Lambda$-doublet transitions can be precisely measured, such molecules are leading candidates for searches for the variation of fine-structure constant and electron to proton mass 
ratio~\cite{koslov2009}.

Nitric oxide is a $\Lambda$-doublet molecule, and has been extensively studied for its numerous applications across physics and chemistry. NO is an atmospheric constituent, which plays a large role in the chemistry and heat budget of the thermosphere, and at lower altitudes is a catalyst in depleting atmospheric ozone~\cite{Campbell2012}.
Because NO is also a neurotransmitter and neurotoxic~\cite{Thomas2008,Korneev2005},  there is practical interest in NO detection. 
A recent promising technique for detection of NO concentration with high sensitivity is through \ry excitation in NO~\cite{Schmidt2018}.

\begin{figure}[t]
%\centering
\includegraphics[scale=1.15,angle=0]{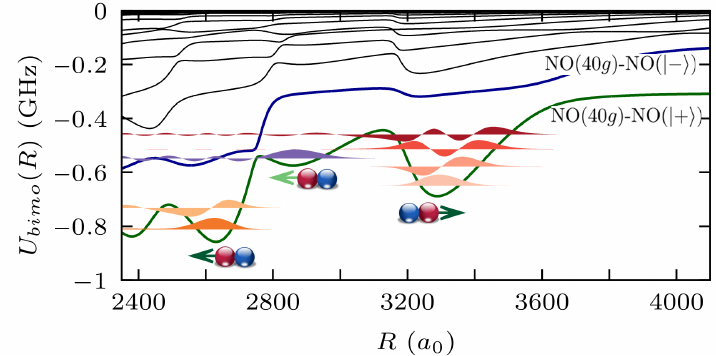}
\caption{$^3\Sigma^+$ 
Born-Oppenheimer potential energy curves near the NO($40g$)-NO(X$^2\Pi_{1/2}$) thresholds. The molecular $\Lambda$-doublet splitting are identified by $|+\rangle$ and $|- \rangle$ labels. The wave functions of long-lived \ry bi-molecular states are shown  at the energy level. The magnitude and direction of orientation of the NO molecular dipole for each electronic well are also shown. All the Hamiltonian terms in~\autoref{eq:fullH} are included in obtaining these potential curves.}
\label{fig:APC:n_40g}
\end{figure}
There are a number of unique features of \ry NO-NO bi-molecular interactions at large distances,  see~\autoref{fig:APC:n_40g}: a) hybridization of nearby $\Lambda$-doublet opposite parity states is efficient even when molecules possess permanent electric dipole moments (PEDM) much below the Fermi-Teller critical dipole, $d_c =1.63$D~\cite{fermi47}, $d_{\text{NO}}\sim 0.16$~D~\cite{gijsbertsen2007,neumann1970}, b) long-lived high electronic angular momentum \ry bi-molecular states ($nf$ and $ng$) with small, but non-zero quantum defects, with extremely large PEDM (a few kilo-Debye) exist, rendering microwave transitions in such molecules possible, and c) highly oriented NO dipoles form which can be coherently controlled for ultracold \ry chemistry, imaging of the molecular structure,
and quantum information processing \cite{Kuznetsova2016}.

The Born-Oppenheimer (BO) Hamiltonian for the \ry NO-NO bi-molecular collision at large distances can be written as 
\begin{equation}
\label{eq:Hamil_adiabatic}
H=H_{\text{Ryd}} +H_{\text{NO}} +H_{\text{e\textendash NO}},
\end{equation}
where $H_{\text{Ryd}}$ is the Hamiltonian for the NO \ry molecule and $H_{\text{NO}}= E_+|+\rangle\langle\, +\,|+E_-|-\rangle\langle -|$ is a two-state $\Lambda$-doublet Hamiltonian for NO, 
with a splitting, 
$E_--E_+=205.95$ MHz~\cite{neumann1970}.

The Hamiltonian $H_{\text{e \textendash NO}}$ contains the \ry electron  and core
interacting with the NO PEDM and 
the scattering of the \ry electron from NO,
\begin{align}
\label{eq:fullH}
H_{\text{e{\textendash}NO}}(\mathbf{R},\mathbf{r})=-\mathbf{d}_{\text{NO}}\cdot \mathbf{F}_e(\mathbf{R},\mathbf{r}) +
\nonumber \\
2\pi a_S(k) \delta\left({\bf r}- {\bf R}\right) 
+ 6 \pi a_P^3(k) \delta\left({\bf r}\right)\overleftarrow{\nabla}\cdot \overrightarrow{\nabla},
\end{align}
where $\mathbf{r}$ and $\mathbf{R}$
are the positions of the \ry electron 
and  NO(X$^2\Pi_{1/2}$) with respect to the core NO$^+$, respectively.
The first term in~\autoref{eq:fullH} describes the interaction of  the NO$^+$ core and the \ry electron 
with the NO(X$^2\Pi_{1/2}$) PEDM $\mathbf{d}_{\text{NO}}$. The internal field due  to the \ry electron and core, $\mathbf{F}_{e}(\mathbf{R},\mathbf{r})=e\left[\frac{\mathbf{r}-\mathbf{R}}{|\mathbf{r}-\mathbf{R}_i|^3}+\frac{\mathbf{R}}{R^3}\right]$, hybridizes the $\Lambda$-doublet states $|+\rangle$  and $|-\rangle$ of NO(X$^2\Pi_{1/2}$).
The total angular momentum excluding spin couples to 
$F_{Z,e}(\mathbf{R},\mathbf{r})$, while the other two perpendicular components of this electric field do not contribute to the interaction. We further assume that the position of NO(X$^2\Pi_{1/2}$) is fixed along the $Z$-axis.

The last two terms in~\autoref{eq:fullH} describe the \ry electron collision with NO.
 These two terms, respectively, reflect the contributions to the low-energy electron-molecule scattering in the $L=0$ ($S$-wave) and $L=1$ ($P$-wave) scattering partial waves~\cite{omont,Sadeghpour2000,Shaffer2018}. The $S$-wave scattering length in the Fermi pseudopotential is, 
 $a_S(k) = -\tan(\delta_S(k))/{k}$, with $\delta_S(k)$ the $L=0$ scattering phase shift, and the  
scattering volume, $a_P^3(k)=-\tan(\delta_P(k))/k^3$, diverges at the position of an $L=1$ scattering  resonance. 
Electron scattering from NO(X$^2\Pi_{1/2}$) can form negative ions, NO$^-($X$^3\Sigma^-)$. 
The electron affinity of NO has been measured by Alle {\it et al.}~\cite{Alle1996}, ${E_A} = 33 \pm 10$~meV. In elastic electron scattering from NO(X$^2\Pi_{1/2}(\nu=0)$), there exists a low-energy resonance at $\sim100$~meV. This resonance has been measured~\cite{zecca1974,Tronc1975} in the channel [$e^-$+NO(X$^2\Pi_{1/2}(\nu=0))\rightarrow$ NO$^-$(X$^3\Sigma(\nu'=1)$)]; there are additional NO$^-$(X$^3\Sigma(\nu'>1)$) resonances, which appear in the inelastic channels. The fundamental vibrational frequency in  NO(X$^2\Pi_{1/2}$) is 
$236$~meV~\cite{OLMAN196462}.

Both NO and NO$^+$ have PEDM, and the resulting dipole-dipole interaction has also been incorporated in the Hamiltonian~\eqref{eq:Hamil_adiabatic}. For the NO$(nl)$-NO bi-molecules, at separations of hundreds a$_0$, these dipolar interactions are several orders of magnitude smaller that the dominant charge-dipole interactions.

The wave function of Hamiltonian~\eqref{eq:Hamil_adiabatic} is expanded in electronic \ry NO and molecular doublet basis and 
the corresponding Schr\"odinger equation,  $H{\bf C} = U_{bimo}(R) {\bf C}$, is solved to obtain the bi-molecular \ry BO potential energy curves $U_{bimo}(R)$, and the electronic wave function ${\bf C}$.

To obtain the wave functions and energies for the \ry NO, a quantum defect description is appropriate. 
The \ry NO BO potential can be written as 
\begin{equation}
    \label{eq:NOBO}
    U_{n\Lambda}(R_{\text{NO}}) = U^+(R_{NO}) - \frac{1}{2(n-\mu_{\Lambda})^2},
    \end{equation}
where $U^+(R_{\text{NO}})$ is the ground-state BO potential curve for NO$^+$(X$^1\Sigma^+$) and   $\Lambda$ is the projection of \ry electron orbital quantum number $l$ on $R_{\text{NO}}$. The quantum defect $\mu_{\Lambda}$  depends on $n$ and $R_{\text{NO}}$.  In our analysis, $R_{\text{NO}}$ is fixed at the equilibrium separation, $R_{\text{NO}}=2.175$ a$_0$. The quantum defects were calculated by Rabad\'an and Tennyson using a $12$ state R-matrix 
method~\cite{Rabadan97}. The Rydberg electron wave functions and energies for different molecular symmetries were obtained by solving the hydrogenic Schr\"odinger equation for non-integer values of $n$~\cite{Bates,Infeld,Kostelecky} using the finite element method.
We note that frame transformation multichannel quantum defect 
theory~\cite{GREENE1985} is a compact formulation of \ry electron interaction with molecular core electrons. In \ry NO, for instance, it results in mixing of Hund's cases (d) and (b) molecular symmetries. This mixing is particularly important at low orbital angular momenta, when the electron spends considerable time near the core and \ry energies will depend on the ion rotational quantum numbers~\cite{Vrakking1996}. Here, we are interested in formation of ultralong-range \ry bi-molecules when the NO molecules are excited into high orbital angular momentum states, $nf$ and  $ng$.

\begin{figure}[t]
%\centering
\includegraphics[scale=1.1,angle=0]{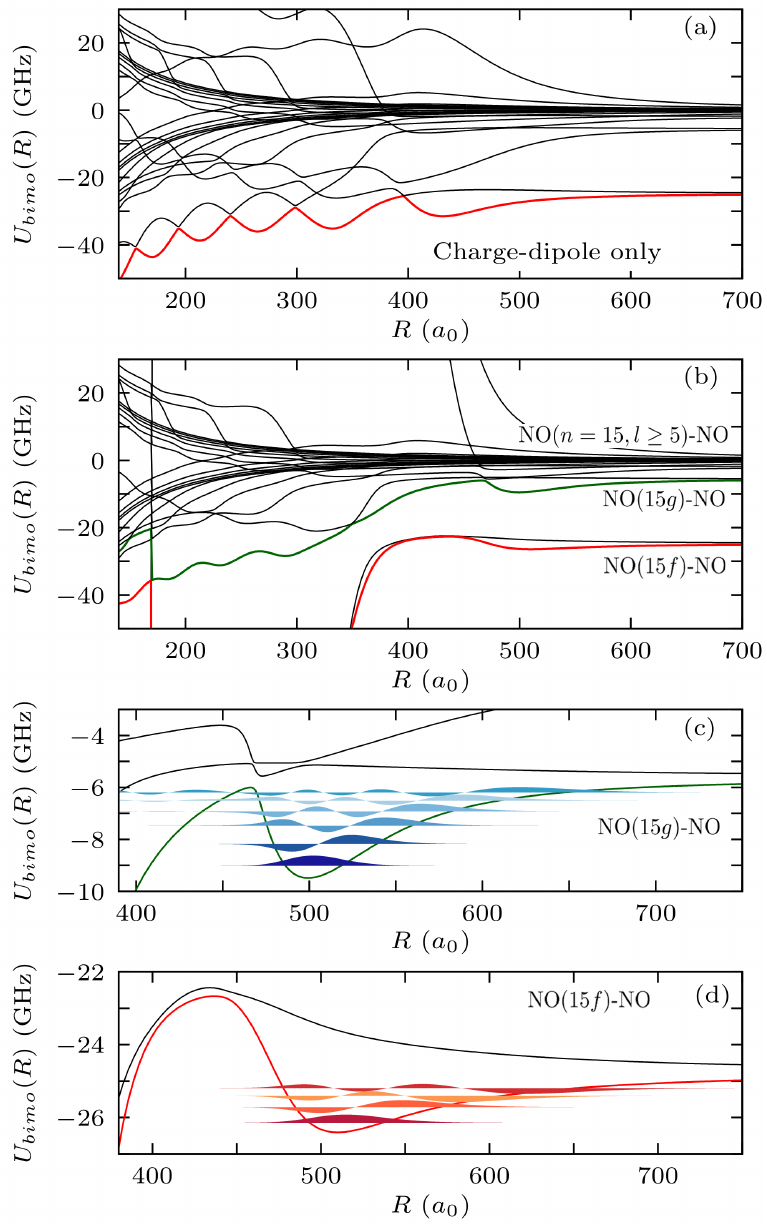}
\caption{
BO potential energy curves of the 
NO($n=15,l\ge 3$)-NO(X$^2\Pi_{1/2}$) ultralong-range Rydberg bi-molecule
including (a) the charge-dipole  only and (b)
all the interactions in the Hamiltonian~\eqref{eq:fullH}.
Details of the curves including all interactions near the (c) NO($15g$)-NO(X$^2\Pi_{1/2}$) and (d) NO($15f$)-NO(X$^2\Pi_{1/2}$) 
dissociation thresholds; the vibrational wave functions are show at the energy level. 
The bound states near the avoided crossing in (c) are quantum reflected states~\cite{Bendkowsk2010}.
}
\label{fig:APC:n_15_all}
\end{figure}

\autoref{fig:APC:n_15_all}~(a) presents the BO potential energy curves when {\it only} the charge-dipole interaction is included in the Hamiltonian~\ref{eq:fullH}~\cite{rittenhouse10,gonzalez15}. 
The near-degenerate $\Lambda$-doublet combined with the small
quantum-defects of the  $nf$ and $ng$ \ry states give rise to several potential wells supporting bound vibrational levels.
By taking into account the low energy scattering of the 
 \ry electron with NO, the complexity of these electronic states is significantly enhanced in~\autoref{fig:APC:n_15_all}~(b).
 Because of the interplay of 
charge--dipole interaction and e$^-$--NO collision, deep  molecular binding occurs at large distances ($R\sim 500$ a$_0$), 
in the exponential tail of NO($15f$) and NO($15g$) \ry wave functions,
see the NO($15f$)-NO(X$^2\Pi_{1/2}$)  and NO($15g$)-NO(X$^2\Pi_{1/2}$)
 BO potential curves in~\autoref{fig:APC:n_15_all}~(c) and~(d), respectively.
 The outer wells support bound levels with 
$200-800$~MHz vibrational spacing. At smaller distances,
$R \lesssim 400$ a$_0$, the $P-$wave resonance introduces avoided crossing and $l-$mixing among the potential energy curves, destroying several inner wells visible 
 in~\autoref{fig:APC:n_15_all}~(a).
 
Because of proximity to hydrogenic degenerate manifolds, 
the $nf$ and $ng$ \ry states have considerable $l$-mixed character,
giving rise to long lifetimes.
For the $nf$ states, lifetimes of $10-50$~ns have been measured~\cite{vraaking1994}, whereas $ng$ states have lifetimes of $0.15 \ \mu \text{s}<\tau < 0.6\ \mu \text{s}$ for $30g < nl< 55g$, against predissociation and vibrational 
autoionization~\cite{Fujii1995}.
Due to the $l$-mixing, the
NO($15f$)-NO(X$^2\Pi_{1/2}$) and NO($15g$)-NO(X$^2\Pi_{1/2}$)  bi-molecules possess fairly large PEDM, $\sim 0.5$~kD.
The rotational constant for the $n=15$ \ry NO bi-molecules is about $0.5$~MHz. As a consequence, microwave spectroscopy of rotational transitions now becomes feasible, and realistic microwave Rabi frequencies of $1-10$~MHz can be achieved.

\begin{figure}[t]
\includegraphics[scale=1.1,angle=0]{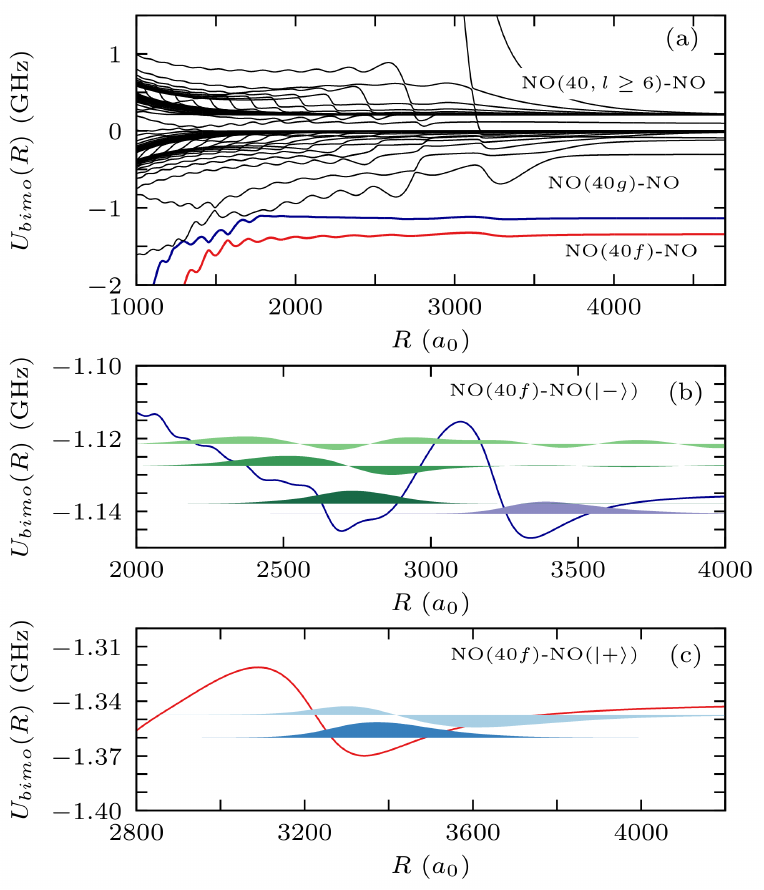}
\caption{(a) BO potential energy curves of the
NO($n=40,l\ge 3$)-NO(X$^2\Pi_{1/2}$) Rydberg bi-molecule
including all the  interactions in the Hamiltonian~\eqref{eq:fullH}.
Details of the curves near the  (b)
NO($40f$)-NO(X$^2\Pi_{1/2}$,$|-\rangle$)
and (c) NO($40f$)-NO(X$^2\Pi_{1/2}$,$|+\rangle$)
dissociation thresholds;
the vibrational wave functions are shown at the energy level.}
\label{fig:APC:n_40_all}
\end{figure}
The BO potential energy curves converging  to the 
NO($40,l\ge3$)-NO(X$^2\Pi_{1/2}$) dissociation threshold,
are shown in~\autoref{fig:APC:n_40_all}. 
Because the $P$-wave resonance affects the energies at shorter distances, its effect on the NO($40f$) and NO($40g$) \ry bi-molecular states is reduced.
The potential curves converging to NO($40f$) 
in Figs.~\ref{fig:APC:n_40_all}~(b) and~(c) support deep vibrational levels. The PEDMs for
NO($40f$)-NO(X$^2\Pi_{1/2}$) and NO($40g$)-NO(X$^2\Pi_{1/2}$)
electronic states are in the $5-7$~kD range. The $\Lambda$-doublet separated upper curve in NO($40f$), for instance, in~\autoref{fig:APC:n_40_all}(b) has a double-well with well-separated molecular states. The right and left wells correspond to NO molecular orientation away and toward the \ry molecular core~\cite{rittenhouse10}. The degree of NO dipole orientation is roughly $40\%$.
This orientation is even larger for the NO($40g$)-NO(X$^2\Pi_{1/2}$) bi-molecules in~\autoref{fig:APC:n_40g}.

The \ry bi-molecules can be experimentally created by excitation of the main optical transition in NO,  A$^2\Sigma^+ (\nu'=0, N'=0) \leftarrow$ X$^2\Pi (\nu=0, N=0)$ at $225$~nm in the 
$\gamma$-band ~\cite{Morrison2008,Schmidt2018,Vrakking1996}. The A$^2\Sigma^+$ state has configuration interaction coefficients,  $94\%$ $(l=0)$, $1\%$ $(l=1)$ and $5\%$ $(l=2)$~\cite{Kaufmann1985}, allowing for excitation of $np$ ($\Delta N =0$), $np$ and $nf$ ($\Delta N=0,\pm 2$) optical transitions. Additional transitions, to H$'(^2\Pi)$ or H$(^2\Sigma^+)$ electronic states, 
and \ry $^2\Sigma^+$ states~\cite{munkes} can produce 
NO$(nf)$-NO and NO$(ng)$-NO \ry bi-molecules. 
In addition, NO$(nf, ng)$-NO \ry bi-molecules can be formed  by excitation 
near the X$^1\Sigma^+$ electronic ground state of NO$^+$ ion with a $328$~nm pulsed laser~\cite{Schmidt2018,Fujii1995}.
In Ref.~\cite{Fujii1995}, the NO$(4f)$ state is an intermediary state to NO$(ng)$ \ry levels.

In summary, we describe the formation of a new class of ultralong-range molecules, comprised of the interaction and collision of a \ry molecule (here NO) with another ground state $\Lambda$-doublet (NO) molecule. Long-lived \ry bi-molecular states with enormous PEDM are predicted. Because the rotational constant of such \ry bi-molecules are in the $0.1-1$~MHz range, microwave spectroscopy of rotational transitions are now within the realm of possibility with realistic microwave Rabi frequencies. 
Cold and ultracold NO samples could be made in a supersonic beam expansion~\cite{Morrison2008,Deller2020} and via 
Sisyphus cooling techniques~\cite{Zeppenfeld2012}, respectively.
In a thermal molecular gas, rotational states are Boltzmann distributed, but yet it is possible for the main rotational transition to be observed. Because the \ry electron couples to the spin symmetries in the molecule, the interaction terms in~\autoref{eq:fullH} can be spin-mixed and BO potentials with mixed spin and orbital angular momenta emerge, which can be manipulated and controlled with small magnetic fields. With a proliferation of activities in trapping of doublet molecules, in particular in optical tweezers~\cite{Anderegg1156} and magnetic traps~\cite{sawyer2007}, \ry excitation in such molecules in cold and ultracold settings is within reach. The ultralong-range \ry bi-molecules predicted in this work hence open a new vista into  cold bi-molecular collisions and control interactions.

R.G.F. acknowledges  financial support  of the  Subprograma Estatal de Movilidad of the 
Programa Estatal de
Promoci\'on del Talento y su Empleabilidad en I+D+i (MECD), the Spanish Project  FIS2017-89349-P (MINECO), 
and the Andalusian  research  group FQM-207.
R.G.F. did this work as a Fulbright fellow at ITAMP 
at Harvard  University. We wish to thank
Seth Rittenhouse
for helpful suggestions. HRS acknowledges a visit to University of Stuttgart as part of the GiRyd program, where the initial discussions were carried out. 

\bibliographystyle{apsrev4-1} 
\bibliography{TriMol}
\end{document}